\documentclass[aps,pra,showpacs,twocolumn]{revtex4-1} 	

\usepackage{graphicx, amsmath, amssymb}
\usepackage{bbold}		
\usepackage{hyperref}

\renewcommand{\Re}{\operatorname{Re}}

\begin{document}

\title{Hyperfine-frequency shifts of alkali-metal atoms during long-range collisions}
\author{B. H. McGuyer}
\affiliation{Department of Physics, Columbia University, 538 West 120th Street, New York, New York 10027-5255, USA}
\date{\today}

\begin{abstract}
Collisions with chemically inert atoms or molecules change the hyperfine coupling $A \, \mathbf{I} \cdot \mathbf{S}$ of an alkali-metal atom through the hyperfine-shift interaction $\delta A \, {\mathbf I} \cdot {\mathbf S}$.  
This interaction is responsible for the pressure shifts of the microwave resonances of alkali-metal atoms in buffer gases, is an important spin interaction in alkali-metal--noble-gas van der Waals molecules, and is anticipated to enable the magnetoassociation of ultracold molecules such as RbSr.  
An improved estimate is presented for the long-range asymptote of this interaction for Na, K, Rb, and Cs.  
To test the results, the change in hyperfine coupling due to a static electric field is estimated and reasonable agreement is found.  
\end{abstract}


\maketitle


During a collision with a chemically inert atom or molecule, the hyperfine coupling 
$A \, \mathbf{I} \cdot \mathbf{S}$ 
between the nuclear spin $\mathbf{I}$ and electronic spin $\mathbf{S}$ of a ground-state alkali-metal atom is altered by the hyperfine-shift interaction,  
\begin{align}\label{ch3:hfs}
H_\text{hfs} = \delta A \, {\mathbf I} \cdot {\mathbf S}, 
\end{align}  
in addition to smaller anisotropic interactions.  
This interaction is responsible for nearly all of the pressure shifts of the microwave resonant frequencies of alkali-metal atoms in cells with buffer gas, which are used in atomic frequency standards (or clocks) and magnetometers~\cite{vanier:1989}.  
This interaction is also important to the study of alkali-metal--noble-gas van der Waals molecules~\cite{gong:2008, mcguyer:2011pra}, 
and is anticipated to enable the formation of certain ultracold molecules, such as RbSr, by magnetoassociation~\cite{zuchowski:2010}.  

The shift parameter $\delta A = \delta A(R)$ in (\ref{ch3:hfs}) is a potential that depends on the colliding pair and their internuclear separation $R$.  
Despite a good amount of theoretical and experimental attention, not much is known yet about the hyperfine-shift potential $\delta A(R)$, especially at small separations $R$.  
Theoretical calculation of $\delta A(R)$ is difficult even for H \cite{rao:1970, ray:1975}, and for the heavy alkali metals is a hard 
problem \cite{camparo:2007jcp}.  
Even at large separations $R$, previous estimates for $\delta A(R)$ disagree by almost a factor of 2~\cite{margenau:1959, adrian:1960, margenau:1961, vanier:1989}.  
The purpose of this Brief Report 
is to provide an improved estimate of the large-$R$ asymptote of $\delta A(R)$ for Na, K, Rb, and Cs.  

Consider a colliding pair with an interaction (or interatomic) potential $V(R)$ that has the asymptotic form 
\begin{align}\label{ch3:c6}
V(R) \approx { - C_6 }{ R^{-6} }
\end{align}
for large separations $R$ where retardation~\cite{casimir:1948} is negligible.  
As derived below, the hyperfine-shift potential $\delta A(R)$ will then have the asymptotic form 
\begin{align}\label{ch3:dAlargeR}
\delta A(R) &\approx  {- \delta A_6} \, {R^{-6}} 
\end{align}
for the same range of $R$.  
The coefficient $\delta A_6$ in (\ref{ch3:dAlargeR}) is related to the magnetic-dipole coupling coefficient $A$ and the van der Waals dispersion coefficient $C_6$ in (\ref{ch3:c6}) by
\begin{align}\label{ch3:A6}
\frac{\delta A_6}{A} \approx \left( \frac{ 2 }{ E_\text{a}} + \frac{1}{E_\text{ab}} \right) C_6,  
\end{align}
where the characteristic energy $E_\text{a}$ depends only on the alkali-metal atom, but 
$E_\text{ab}$ depends on the colliding pair.  

Previous work has produced expressions of the same form as (\ref{ch3:A6}), but with differing estimates for $E_\text{a}$ and $E_\text{ab}$ \cite{margenau:1961, vanier:1989}.  
Of the two terms in (\ref{ch3:A6}), the second with $E_\text{ab}$ is a small contribution, typically 10\% for noble-gas perturbers in previous work, so the disagreement between the estimates for $E_\text{a}$ is the most significant:  
Vanier and Audoin \cite{vanier:1989} estimate $E_\text{a}$ as a rough average of optical ($D_1$ and $D_2$) transition and ionization energies, while 
Herman and Margenau \cite{margenau:1961} estimate $E_\text{a}$ as the alkali-metal ionization energy $I_\text{a}$ after numerical work.  
As shown in Table~\ref{ch3:tab:2}, these previous estimates differ by roughly a factor of 1.5. 

To provide an improved estimate, let us now derive explicit forms for the energies $E_\text{a}$ and $E_\text{ab}$ in (\ref{ch3:A6}).  
In what follows, we will use less approximation than Ref.~\cite{vanier:1989} and use experimental values and tabulated wave functions that were not available a half century ago with Refs.~\cite{margenau:1961, margenau:1959}.  
Estimates for the error of relation (\ref{ch3:A6}) and for the characteristic energies $E_\text{a}$ for Na, K, Rb, and Cs will be provided. 
To test the results, the values for $E_\text{a}$ are used to estimate the change in hyperfine coupling due to a static electric field.  

Consider an alkali-metal atom at position ${\bf x}_\text{a}$ and a perturbing atom or molecule at position ${\bf x}_\text{b} = {\bf x}_\text{a} + {\bf R}$, both of which are in their ground states.  
For large enough $R = |{\bf R}|$ and ignoring retardation \cite{casimir:1948}, the leading-order interaction $U$ responsible for the $V(R)$ of (\ref{ch3:c6}) is the dispersive van der Waals interaction between the instantaneous electric-dipole moment ${\bf p}_\text{b}$ of the perturber and the electric field ${\bf E}_\text{a}({\bf x}_\text{b})$ from the instantaneous moment ${\bf p}_\text{a}$ of the alkali-metal atom, 
\begin{align} \label{U1}
U = - {\bf p}_\text{b} \cdot {\bf E}_\text{a}({\bf x}_\text{b}). 
\end{align}
The moment ${\bf p}_\text{b} = - |e| {\bf r}_\text{b}$, where $e$ is the electronic charge and ${\bf r}_\text{b}$ is the sum of all the positions relative to ${\bf x}_\text{b}$ of the electrons of the perturber.  
The field ${\bf E}_\text{a}({\bf x}_\text{b}) = {\bf p}_\text{a} \cdot (3 {\bf n} {\bf n} - \mathbb{1})/R^3$, where ${\bf n} = {\bf R}/R$ and $\mathbb{1}$ is the identity dyadic tensor.  
The moment ${\bf p}_\text{a} = - |e| {\bf r}_\text{a}$, where ${\bf r}_\text{a} = {\bf r} + {\bf r}_\text{c}$ is the sum of the positions ${\bf r}$ and ${\bf r}_\text{c}$ relative to ${\bf x}_\text{a}$ of the alkali-metal single valence and core electrons, respectively.  
Thus we may write (\ref{U1}) as 
\begin{align} \label{U2}
U &= e^2 ( {\bf r}_\text{a} \cdot {\bf r}_\text{b} - 3 \, z_\text{a} z_\text{b} ) / R^3, 
\end{align} 
where $z_\text{a} = {\bf r}_\text{a} \cdot {\bf n}$ and $z_\text{b} = {\bf r}_\text{b} \cdot {\bf n}$.

\begin{table}[t!] 
\caption{\label{ch3:tab:2}
Characteristic energies $E_\text{a}$ (eV) of expression (\ref{ch3:A6}) estimated using (\ref{ch3:CalcEa}). 
Previously suggested values for $E_\text{a}$ are included for comparison:  
$\overline{E}_e$ from Vanier {\it et al.}~\cite{vanier:1989} 
and the alkali-metal ionization energy $I_\text{a}$~\cite{nistspectra} from Herman {\it et al.}~\cite{margenau:1961}.  
The values in parentheses are uncertainties in the last digits. 
}
\begin{ruledtabular}
\begin{tabular}{l | c c c c }
Alkali metal: 	& Na				& K 				& Rb 			& Cs \\
\hline
$E_\text{a}$ (this work): 
			& 6.55(33)		& 5.31(28)		& 5.05(37)		& 4.59(48) \\		
Vanier {\it et al.}~\cite{vanier:1989}:  
			& 3.62	& 2.98	& 2.88 	& 2.66 \\
Herman {\it et al.}~\cite{margenau:1961}:  
			& 5.14	& 4.34	& 4.18 	& 3.89\\
\end{tabular}	
\end{ruledtabular}
\end{table}

Following Adrian \cite{adrian:1960}, let us treat both $U$ and the contact magnetic-dipole hyperfine interaction for the alkali-metal valence electron \cite{arimondo:1977}, 
\begin{align}\label{ch3:Hhf}
H_\text{hf} = \frac{ 8 \pi }{ 3 } g_S \mu_B   \frac{ \mu_I }{ I } \delta({\bf r})\, {\bf I} \cdot {\bf S}, 
\end{align}
as simultaneous perturbations to the colliding pair.  
Let $|\mu \nu \rangle$ denote the tensor product of the unperturbed wave functions for the the $\mu$th eigenstate of the alkali-metal atom 
and the $\nu$th eigenstate of the perturber. 
Let $E_{\mu\nu}$ denote the energy of this state, and let $\mu = 0$ and $\nu = 0$ denote ground states.   

Let us assume that the ground state of the perturber is spherically symmetric, like the alkali-metal ground $S$ state, such that $\langle 00|U|\mu\nu\rangle = 0$ if either $\mu$ or $\nu = 0$.  
Then the first-order perturbation to the total ground-state energy $E_{00}$ is 
\begin{align} 	\label{dE00,1}
\delta E_{00,1} = \langle 00 | H_\text{hf} | 00 \rangle = A \langle {\bf I} \cdot {\bf S} \rangle, 
\end{align}
where $A$ is the magnetic-dipole coupling coefficient of the unperturbed alkali-metal atom. 
Here and subsequently, angle brackets denote ground-state expectation values.  

The second-order perturbation $\delta E_{00,2}$ contains the long-range van der Waals interaction 
\begin{align} 	\label{dE00,2}
\delta E_{00,2}^\text{(vdW)} = \sum_{\mu,\nu \neq 0} \frac{|\langle 0 0| U | \mu \nu \rangle|^2}{E_{00}-E_{\mu\nu}} = - C_6 R^{-6},  
\end{align}
which may be expressed in more standard forms \cite{stanton:1994}, as well as second-order hyperfine terms, but 
no cross terms because $\langle 0 0 | H_\text{hf} | \mu \nu \rangle = 0$ for $\nu \neq 0$.  

The leading-order hyperfine-shift interaction (\ref{ch3:hfs}) comes from terms in the third-order perturbation $\delta E_{00,3}$ that are linear in $H_\text{hf}$, 
\begin{align} 	\label{dE00,3}
\delta E_{00,3}^\text{(lin)} =& \, 2 \, \Re \sum_{\eta,\mu,\nu \neq 0} \frac{ \langle 0 0| H_\text{hf} | \eta 0 \rangle \langle \eta 0 | U | \mu \nu \rangle \langle \mu \nu | U | 00 \rangle}{(E_{00}-E_{\eta 0})(E_{00}-E_{\mu\nu})} \nonumber \\
	& - \langle 00 | H_\text{hf} | 00 \rangle \sum_{\rho,\sigma \neq 0} \frac{|\langle 0 0| U | \rho \sigma \rangle|^2}{(E_{00}-E_{\rho\sigma})^2}.     
\end{align}
Note that an additional linear term is zero because 
$\langle \mu \nu | H_\text{hf} | \rho \sigma\rangle = 0$ unless both $\mu$ and $\rho$ are spherically symmetric $S$ states, in which case $\langle 0 0 | U | \mu \nu \rangle = 0$. 

To proceed further, let us make two changes.  
First, approximate $(E_{00}-E_{\mu\nu}) \approx - E_\text{ab}$ in the first term and $(E_{00}-E_{\rho\sigma})^2 \approx - E_\text{ab} (E_{00}-E_{\rho\sigma})$ in the second term of (\ref{dE00,3}). 
Second, use closure to remove the sums over $\mu$ and $\nu$ in the first term. 
Lacking explicit knowledge of the perturber, a reasonable choice is to define $E_\text{ab}$ so that these two changes return (\ref{dE00,2}) to itself, which gives 
\begin{align}	\label{Eabv1}
E_\text{ab} = -\frac{\langle 00| U^2 |00\rangle}{\delta E_{00,2}^\text{(vdW)}} \approx \frac{2 e^4 \langle r_\text{a}^2 \rangle\langle r_\text{b}^2\rangle}{3 \,C_6},
\end{align}
where $r_\text{a} = |{\bf r}_\text{a}|$ and $r_\text{b} = |{\bf r}_\text{b}|$. 
The approximation on the right assumes uncorrelated electronic positions. 
For noble-gas perturbers, one can show that the $E_\text{ab}$ of (\ref{Eabv1}) are larger than the previous estimates of Refs.~\cite{vanier:1989, margenau:1961}.  



With these changes, (\ref{dE00,3}) simplifies to the form 
\begin{align} 	\label{dE00,3approx}
\delta E_{00,3}^\text{(lin)} \approx& \; \delta E_{00,1}^{\vphantom{W}} \delta E_{00,2}^\text{(vdW)} \left( \frac{2}{E_\text{a}} + \frac{1}{E_\text{ab}} \right)
\end{align}
where the characteristic energy 
\begin{align} 	\label{Eav1}
\frac{1}{E_\text{a}} =& \sum_{\eta \neq 0} \frac{ \langle 0 0| H_\text{hf} | \eta 0 \rangle \langle \eta 0 | U^2 | 00 \rangle}{(E_{00}-E_{\eta 0})\langle 00 | H_\text{hf} | 00 \rangle \langle 00 | U^2 | 00 \rangle}. 
\end{align}
Relation (\ref{ch3:A6}) follows from using (\ref{dE00,3approx}) with (\ref{dE00,1}) and (\ref{dE00,2}).   

To numerically estimate $E_\text{a}$, let us simplify (\ref{Eav1}) as follows. 
First, let us ignore the alkali-metal core electrons, since the single valence electron is the dominant contributor to the interaction (\ref{U2}).
Second, restrict the sum over $\eta$ to alkali-metal $S$ states, since the contact interaction (\ref{ch3:Hhf}) is nonzero only for these states. 

Let $|n\rangle$ denote the $S$ state wave function for an unperturbed alkali-metal valence electron with principle quantum number $n$ and energy $E_{n}$. 
Let $|g\rangle$ denote the ground $S$ state with energy $E_g$ and $n=g=$ 3, 4, 5, and 6 for Na, K, Rb, and Cs, respectively.  
Note that each $S$ state $|n\rangle$ has a magnetic-dipole coupling coefficient 
\begin{align}\label{ch3:An}
A_n = \frac{ 8 \pi }{ 3 } g_S \mu_B   \frac{ \mu_I }{ I } |\psi_n(0)|^2,
\end{align}
with the free-atom $A_g = A$, and where $|\psi_n(0)|^2$ is the valence-electron probability density at the nucleus.  

Using this notation with the square-root formula $\langle g | H_\text{hf} | n \rangle = \sqrt{\langle n | H_\text{hf} | n \rangle \langle g | H_\text{hf} | g \rangle}$, 
we may approximate (\ref{Eav1}) as  
\begin{align}\label{ch3:CalcEa}
\frac{1}{E_\text{a}} \approx \frac{ 1 }{ \langle r^2 \rangle } \sum_{n > g} \frac{ \langle g | r^2 | n \rangle \sqrt{ |A_n / A_g| } }{ E_g  - E_n }, 
\end{align}
where $r = |{\bf r}|$. 
The values of $E_\text{a}$ in Table I were numerically estimated using this as described below.  
Note that the square-root formula is expected to remain accurate to better than 1\% when relativistic and many-body effects are included \cite{bouchiat:1988,flambaum:2000}.  
This expression for $E_\text{a}$ may be derived more concisely using an effective electric-dipole polarizability for the perturber \cite{mcguyer:thesis}.  
However, such an approach neglects the smaller term with $E_\text{ab}$ in (\ref{ch3:A6}). 

Before we continue, let us address the accuracy of these results.  
To compute $\delta A_6 / A$, one may use the relation (\ref{ch3:A6}) with a value of $E_\text{a}$ from Table I, of $E_\text{ab}$ estimated using (\ref{Eabv1}), and of $C_6$ fot the colliding pair, many estimates of which are available in Refs.~\cite{mitroy:2007:I,mitroy:2007,derevianko:2010}. 
Overall, the error of (\ref{ch3:A6}) is most likely dominated by the error of the approximations used to derive the simplified form (\ref{ch3:CalcEa}) for $E_\text{a}$, in particular, the neglect of the alkali-metal core electrons.  
An estimate for this error is the fractional contribution of the alkali-metal core electrons to $C_6$, which is roughly 20\% for alkali-metal--noble-gas pairs \cite{mitroy:2007}.  
Compared to this error, one can show that the contribution of the $E_\text{ab}$ of (\ref{Eabv1}) to (\ref{ch3:A6}) is often negligible.  
As a result, only $E_\text{a}$ is estimated below.  
More rigorous work is required for accuracy beyond this level. 

Following Oreto {\it et al.}~\cite{oreto:2004}, 
let us write the wave function for the alkali-metal valence $S$ state $|n\rangle$ as 
\begin{align}
\psi_{n,m}({\bf r}, \sigma) = \frac{ P_{n0}(r) }{ 2 \sqrt{\pi} \, r} \delta_{\sigma m}, 
\end{align}
where $m$ is the azimuthal quantum number, the electronic spin variable $\sigma = \pm 1/2$, and $P_{n0}(r)$ is the (real-valued) radial wave function.  
Then the remaining matrix elements in (\ref{ch3:CalcEa}) simplify to radial integrals, 
\begin{align}\label{ch3:gSr2nS}
\langle g | r^2 | n \rangle = \int_0^\infty P_{g0}(r) P_{n0}(r)r^2dr.  
\end{align}
For the ground-state functions $P_{g0}(r)$ and expectations $\langle r^2 \rangle$, the tabulated Roothaan-Hartree-Fock (RHF) wave functions and values of Bunge {\it et al.}~\cite{bunge:1993} were used for Na, K, and Rb, and those of McLean and McLean \cite{mclean:1981} (triple-zeta-valence form) were used for Cs.  

Coulomb-approximation (CA) wave functions were used for the excited-state functions $P_{n0}(r)$. 
Following Oreto {\it et al.}~\cite{oreto:2004}, the CA functions are given by the asymptotic series
\begin{align}\label{ch3:CA}
P_{n0}(r) = \sum_{q=0}^p c_q e^{-r/n^*}r^{n^*-q}, 
\end{align}
where the effective quantum number 
$n^* = \sqrt{ R_\infty / ( I_\text{a} - E_{n} )}$, 
$R_\infty$ is the Rydberg constant, and $I_\text{a}$ is the alkali-metal ionization energy.  
Up to overall normalization, the coefficients $c_q$ are given by the recurrence relation
$c_q / c_{q-1} = n^* (n^* - q)(n^* - q + 1) / ( 2 q)$.
The upper limit $p$ of the series (\ref{ch3:CA}) was chosen to give the best convergence at $r = 1$~Bohr.  The CA functions were normalized such that 
\begin{align}
\int_{0.1}^\infty P_{n0}(r)^2dr = 1, 
\end{align}
where the lower bound is 0.1~Bohr.  
To match the RHF and square-root formula conventions, the signs of the CA functions were chosen so that the nuclear values of the unapproximated functions $P_{n0}'(0)$ are positive, using  
\begin{align}
\frac{P_{n0}'(0)}{|P_{n0}'(0)|} = (-1)^{n+g} \lim_{r \rightarrow \infty} \frac{P_{n0}(r)}{|P_{n0}(r)|}, 
\end{align}
which lead to negative values for all the elements (\ref{ch3:gSr2nS}). 

One consequence of using RHF functions for the ground state and CA functions for the excited states is that the combined set of radial functions is not perfectly orthogonal.  That is, the numerical integrals 
\begin{align}
\int_{0.1}^\infty P_{g0}(r) P_{n0}(r) dr 
\end{align}
are not exactly zero for $n \neq g$, but are, for example, between 0.02 and 0.13 for Na--Cs with the worst case of $n = g + 1$.  
However, the orthogonality quickly improves with $n$, as the accuracy of the CA functions improves with $n$.  
This error is partially suppressed because it is due to the inaccuracy of the CA functions near the nucleus, where the operator $r^2$ in (\ref{ch3:gSr2nS}) contributes least.  

For the physical parameters in (\ref{ch3:CalcEa}), experimental values were used where available, and extrapolated otherwise.  
For the coupling coefficients $A_n$, the values for $^{23}$Na, $^{39}$K, $^{87}$Rb, and $^{133}$Cs from Arimondo {\it et al.}~\cite{arimondo:1977} and Sansonetti~\cite{sansonetti:2006Rb, sansonetti:2008K, sansonetti:2008Na, sansonetti:2009} were used. 
Though the $A_n$ are isotope dependent, the ratio $A_n/A_g$ is expected to be isotope independent to better than 1\%~\cite{persson:2013, arimondo:1977}.  
For the energies $I_\text{a}$ and $E_{n}$, the values from Ref.~\cite{nistspectra} were used.  

In general, the parameters $A_n$ are the least available.  
The extrapolation of $A_n$ to higher $n$ used a linear fit to a plot of $\ln(A_n)$ vs $\ln(n^*)$, which is very nearly a straight line with a slope of almost exactly $-$3, in agreement with semiempirical formulas for $A_n$ \cite{arimondo:1977}.  
The extrapolation of $n^*$ and $E_{n}$ used a linear fit to a plot of $n^*$ vs $n$, which is very nearly a straight line, in agreement with semiempirical formulas using a quantum defect \cite{happer:book}.  
The matrix elements (\ref{ch3:gSr2nS}) were explicitly calculated up to $n=35$.  
The limit $p$ for $P_{n0}(r)$ was optimized where $E_{n}$ is available, and extrapolated to higher $n$ by noticing that $p$ is very nearly equal to $n$ at large $n$, up to a constant offset.  
The matrix elements (\ref{ch3:gSr2nS}) were extrapolated to higher $n$ using a linear fit to the large-$n^*$ asymptote of a plot of $\ln(-\langle g | r^2 | n \rangle)$ vs $\ln(n^*)$, in the region $n$ = 30--35, which gave intercepts and slopes close to 3.5 and $-$1.5, respectively, for each alkali-metal atom.  
Such a dependence is expected because as $n$ increases, the CA functions $P_{n0}(r)$ converge to the same shape over the important range of $r$, up to normalization.  

Using the values and extrapolations described above, expression (\ref{ch3:CalcEa}) was summed to $n=500$, a limit large enough to approximate including all $n$.  
The sums converged quickly, with the highest terms contributing at least 1\% being $n$ = 15, 15, 17, and 18 for Na, K, Rb, and Cs.  
The extrapolation for $n>35$ contributed roughly $-6$\%, $-3$\%, $-2$\%, and $-1$\% for Na, K, Rb, and Cs.  
As Table~\ref{ch3:tab:2} shows, the resulting values of $E_\text{a}$ are significantly larger than those from previous work.  
To test these values, they are used in the Appendix to estimate the change in hyperfine coupling due to a static electric field, and reasonable agreement is found.  
These values allow for an improved estimate of the asymptotic form (\ref{ch3:dAlargeR}) of $\delta A(R)$ using (\ref{ch3:A6}). 

The uncertainties given for the $E_\text{a}$ in Table~\ref{ch3:tab:2} include those of the square-root formula, the experimental values, the isotope dependence of $A_n/A_g$, and the extrapolations, as well as the estimated effects of the radial integration bounds and the CA--RHF non-orthogonality.  
Except for Na, for which the extrapolation of $p$ was significant, the CA--RHF non-orthogonality was the dominant contributor.  
The uncertainties do not include estimated errors for the derivation of (\ref{ch3:CalcEa}) or its use in (\ref{ch3:A6}), which were discussed earlier. 
Finally, note that the $E_\text{a}$ are isotope independent within the uncertainties given.  

In summary, an improved estimate has been provided for the long-range hyperfine-shift interaction (\ref{ch3:hfs}) of Na, K, Rb, and Cs.  
Future work is required to further elucidate the poorly known hyperfine-shift potential $\delta A(R)$.

\begin{acknowledgments}
I am grateful to W.~Happer for many helpful discussions and for suggesting this project. The majority of this work was performed at Princeton University and supported by the Air Force Office of Scientific Research.  
\end{acknowledgments}

\appendix*
\section{Hyperfine-frequency shifts of alkali-metal atoms from static electric fields}\label{append}

To test the values estimated above for the $E_\text{a}$ of (\ref{ch3:CalcEa}), let us use them to estimate the scalar, static Stark shifts of the hyperfine couplings of $^{23}$Na, $^{39}$K, $^{87}$Rb, and $^{133}$Cs. 
%
Consider a ground-state alkali-metal atom in the presence of a uniform, static electric field ${\bf E}_\text{z}$ along the Cartesian unit vector ${\bf z}$. 
Similar to before, the hyperfine coupling is altered by the hyperfine-shift interaction (\ref{ch3:hfs}), in addition to smaller anisotropic interactions (or tensor Stark shifts) \cite{mitroy:2010, mowat:1972}.  
However, the parameter $\delta A$ is to leading order proportional to $|{\bf E}_\text{z}|^2$.  
The hyperfine transition frequency $\nu = A (I + 1/2)/h$ is shifted by  
\begin{align}	\label{k}
\delta \nu = \delta A (I + 1/2)/h \approx  k |{\bf E}_z|^2, 
\end{align}
where $h$ is the Planck constant and $I$ is the nuclear spin quantum number.  
The isotope-dependent Stark-shift coefficients $k$ are known very precisely for several alkali-metal atoms, in part, because they characterize the black-body radiation shift in microwave atomic clocks \cite{mitroy:2010}. 

The dominant interaction responsible for the Stark shift is that of the instantaneous moment ${\bf p}_\text{a} = - |e| {\bf r}$ of the alkali-metal valence electron with the field ${\bf E}_\text{z}$, $U' = - {\bf p}_\text{a} \cdot {\bf E}_z = |e {\bf E}_\text{z}| z$, where $z = {\bf r} \cdot {\bf z}$. 
To estimate $k$, let us treat both $U'$ and the $H_\text{hf}$ of (\ref{ch3:Hhf})  as simultaneous perturbations.  
Then the first-order perturbation to the ground-state energy $E_g$ is $\delta E_{g,1} = \langle g | U' + H_\text{hf}| g \rangle =  A \langle {\bf I} \cdot {\bf S} \rangle$. 
The second-order perturbation $\delta E_{g,2}$ contains a common-mode Stark shift, $\delta E_{g,2}^{\text{(Stark)}} = - \alpha_\text{a}(0) | {\bf E}_z |^2/2$, as well as second-order hyperfine-interaction terms.   
The static polarizability is the standard result \cite{vanier:1989, bonin:1997, mitroy:2010} 
\begin{align}	\label{StandardAlpha}
\alpha_\text{a}(0) = 2 \sum_{\mu \neq g}  \frac{ |\langle g | U' | \mu \rangle|^2}{E_\mu - E_g} = \frac{2 e^2}{3} \sum_{\mu \neq g}  \frac{\langle g | {\bf r} | \mu \rangle \cdot \langle \mu | {\bf r} | g \rangle}{E_\mu - E_g}.
\end{align}
Here and subsequently, a sum over $\mu$ denotes a sum over all excited states $|\mu\rangle$ with energies $E_\mu$ and the continuum. 

The leading-order shift (\ref{k}) comes from the terms in the third-order perturbation $\delta E_{g,3}$ that are linear in $H_\text{hf}$, 
\begin{align}	\label{Eg3p}
\delta E_{g,3}^{\text{(lin)}} =& \, 2 \, \Re \sum_{\mu, n \neq g} \frac{\langle g | H_\text{hf} | n \rangle \langle n | U' | \mu \rangle \langle \mu | U' | g \rangle}{(E_g - E_n)(E_g - E_\mu)}  \nonumber \\
	&- \langle g | H_\text{hf} | g \rangle \sum_{\mu \neq g} \frac{ |\langle g | U' | \mu \rangle |^2}{(E_\text{g} - E_\mu)^2}. 
\end{align}
Note that an effective polarizability operator \cite{bonin:1997} allows a more concise derivation of the first term, but neglects the second term above. 
Both terms are nearly equal, so such an approach underestimates $k$ by roughly half.

\begin{table}[b] 
\caption{\label{Tab2}
Stark-shift coefficients $k$ ($10^{-10}$ Hz/(V/m)$^{2}$) of expression (\ref{k}) estimated using (\ref{kapprox}). 
Experimental values from Mitroy {\it et al.}~\cite{mitroy:2010} are included for comparison. 
The values in parentheses are uncertainties in the last digits, which for this work only include the contributions due to $E_\text{a}$. 
}
\begin{ruledtabular}
\begin{tabular}{l | c c c c }
Alkali metal: 	& $^{23}$Na		& $^{39}$K 		& $^{87}$Rb 		& $^{133}$Cs \\
\hline
$k$ (this work): 
			& $-$0.120(3)		& $-$0.074(2)		& $-$1.29(4) 		& $-$2.50(13) \\
$k$ (expt., \cite{mitroy:2010}): 
			& $-$0.124(3)		& $-$0.071(2)		& $-$1.23(3) 		& $-$2.271(4) 
\end{tabular}	
\end{ruledtabular}
\end{table}

We can recover the form (\ref{ch3:CalcEa}) for $E_\text{a}$ in (\ref{Eg3p}) with the following two changes. 
First, substitute $(E_g - E_\mu) \approx - E_D$ and $(E_g - E_\mu)^2 \approx E_D (E_\mu - E_g)$ in the denominators, where $E_D$ is a weighted average of the $D_1$ and $D_2$ transition energies to the first excited $P$ state. 
This substitution is accurate to within 5\% in calculating $\alpha_\text{a}(0)$ with (\ref{StandardAlpha})  (e.g., using Ref.~\cite{safronova:1999}). 
Second, use closure and $\langle g | U' | g \rangle = 0$ to remove the first sum over $\mu$. 


Noting $ \delta A \approx \delta E_{g,3}^{\text{(lin)}}/\langle {\bf I} \cdot {\bf S} \rangle$ and using (\ref{Eg3p}) with the changes described above, the square-root formula, (\ref{ch3:CalcEa}), and (\ref{StandardAlpha}), 
we find that the $k$ in (\ref{k}) is approximately 
\begin{align}	\label{kapprox}
k \approx - \frac{A (I + 1/2)}{2 h \, E_D} \left( \frac{4 e^2 \langle r^2 \rangle}{3 \, E_\text{a}} + \alpha_\text{a}(0) \right).  
\end{align}
Using this with the $E_\text{a}$ of Table I, the values of $k$ in Table II were numerically estimated for $^{23}$Na, $^{39}$K, $^{87}$Rb, each with $I=3/2$, and $^{133}$Cs with $I=7/2$.  
RHF $\langle r^2 \rangle$ 
and experimental $A$ 
were used as before.  
The first excited $P$ state term energies from Ref.~\cite{nistspectra} were used for $E_D$, and experimental values from Mitroy {\it et al.}~\cite{mitroy:2010} for $\alpha_\text{a}(0)$. 

The uncertainties for $k$ in Table II are solely due to those for $E_\text{a}$ in Table I. 
As shown, the estimates of $k$ agree with measurements to within twice this uncertainty. 


%

\end{document}